\documentclass[12pt,a4paper, reprint, aip, jap, twocolumn, amsmath, amssymb, aps]{revtex4-1} 
\usepackage{graphicx}
\usepackage{floatrow} 
\usepackage[paperwidth=210mm,paperheight=297mm,centering,hmargin=2cm,vmargin=2.1cm]{geometry}
\usepackage{mathptmx}      %% use fitting times fonts also in formulas
\usepackage{amsmath}
\usepackage{setspace}
\usepackage{natbib}
\begin{document}

\title{Spin current generation from sputtered Y$_3$Fe$_5$O$_{12}$ films}

\author{J. Lustikova}
 \email{lustikova@imr.tohoku.ac.jp}
 \affiliation{Institute for Materials Research, Tohoku University, Sendai 980-8577, Japan}

\author{Y. Shiomi}
 \affiliation{Institute for Materials Research, Tohoku University, Sendai 980-8577, Japan}

\author{Z. Qiu}
 \affiliation{WPI Advanced Institute for Materials Research, Tohoku University, Sendai 980-8577, Japan}

\author{T. Kikkawa}
 \affiliation{Institute for Materials Research, Tohoku University, Sendai 980-8577, Japan}

\author{R. Iguchi}
 \affiliation{Institute for Materials Research, Tohoku University, Sendai 980-8577, Japan}

\author{K. Uchida}
 \affiliation{Institute for Materials Research, Tohoku University, Sendai 980-8577, Japan}
 \affiliation{PRESTO, Japan Science and Technology Agency, Saitama 332-0012, Japan}

\author{E. Saitoh}
 \affiliation{Institute for Materials Research, Tohoku University, Sendai 980-8577, Japan}
 \affiliation{WPI Advanced Institute for Materials Research, Tohoku University, Sendai 980-8577, Japan}
 \affiliation{CREST, Japan Science and Technology Agency, Tokyo 102-0076, Japan}
 \affiliation{Advanced Science Research Center, Japan Atomic Energy Agency, Tokai 319-1195, Japan}

\date{\today} 

\begin{abstract}
Spin current injection from sputtered yttrium iron garnet (YIG) films into an adjacent platinum layer has been investigated by means of the spin pumping and the spin Seebeck effects. Films with a thickness of 83 and 96 nanometers were fabricated by on-axis magnetron rf sputtering at room temperature and subsequent post-annealing. 
From the frequency dependence of the ferromagnetic resonance linewidth, the damping constant has been estimated to be $(7.0\pm1.0)\times 10^{-4}$. 
Magnitudes of the spin current generated by the spin pumping and the spin Seebeck effect are of the same order as values for YIG films prepared by liquid phase epitaxy.
The efficient spin current injection can be ascribed to a good YIG$\mid$Pt interface, which is confirmed by the large spin-mixing conductance $(2.0\pm0.2)\times 10^{18}$ m$^{-2}$. 
\end{abstract}

\maketitle

\section{Introduction}
%%%%%%%%%%

Spintronics is an aspiring field of electronics which incorporates the spin degree of freedom into charge-based devices. 
Among the main interests in spintronics are generation, manipulation and detection of spin current, the flow of spin angular momentum. Pure spin current unaccompanied by charge current
has high potential to open a path to
new information technology free from the Joule heating.

\begin{figure}[b]
\centerline{\includegraphics[width=8cm, angle=0, bb=0 0 240 100]{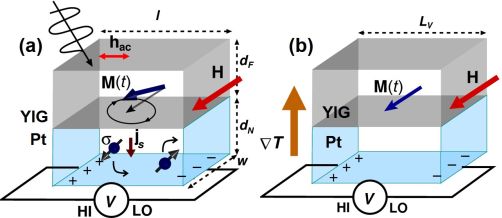}}
\caption{\label{fig1}
Schematic illustrations of the experimental setup for spin pumping (SP) and the spin Seebeck effect (SSE). (a) SP and the inverse spin Hall effect (ISHE). $\mathbf{H}$, $\mathbf{h}_{\text{ac}}$, $\mathbf{M}(t)$, $\mathbf{j}_s$ and $\boldsymbol{\sigma}$ denote the static magnetic field, the microwave magnetic field, the magnetization vector, the direction of spin current generated by SP and the spin-polarization vector of the spin current, respectively. The bent arrows in the Pt layer denote the motion of the electrons under the influence of the spin-orbit coupling which leads to the appearance of a transverse electromotive force (ISHE).
(b) Longitudinal SSE in a YIG$\mid$Pt bilayer film. $\boldsymbol{\nabla} T$ denotes the temperature gradient. Spin current is generated along $\boldsymbol{\nabla} T$ due to SSE and the electromotive force by ISHE in Pt appears in a direction perpendicular both to the sample magnetization and to the temperature gradient. }
\end{figure}

For spin current generation in thin-film systems, two dynamical methods are the spin pumping \cite{silsbee,tserkovnyak-PRL, mizukami, azevedo, saitoh, costache, kajiwara, ando} and the spin Seebeck effect.\cite{uchida2008, uchidanmat2010, jaworski, uchidaAPL, kirihara, kikkawa} In spin pumping [SP, Fig. \ref{fig1}(a)], spin current is generated by magnetization dynamics in the ferromagnet.  
The magnetization vector of a ferromagnet irradiated by a microwave precesses when the ferromagnetic resonance condition is fulfilled. This precession motion relaxes not only by damping processes inside the ferromagnet (F), but also by emission of spin current into the adjacent non-magnetic conductor (N) by exchange interaction at the F$\mid$N interface.\cite{silsbee, tserkovnyak-PRL, mizukami}  
In the spin Seebeck effect (SSE), spin current is generated in the presence of a temperature gradient across the ferromagnet. The simplest setup for SSE is the so-called longitudinal configuration [Fig. \ref{fig1}(b)], where the temperature difference is applied parallel to the direction of spin injection. Given that the ferromagnet is attached to a non-magnetic conductor, spin current is emitted from the ferromagnet into the neighbouring non-magnetic metal by thermal spin pumping.\cite{adachi, rezende}

%%%%%%%%%%%%%%%%%

Notably, these two mechanisms of spin current generation do not require that the ferromagnet be a conductor. The use of an insulator enables generation of pure spin currents 
and limits transport mediated by conduction electrons to the adjacent non-magnetic metal. The ferrimagnet yttrium iron garnet (Y$_3$Fe$_5$O$_{12}$, YIG) is a material of choice as a spin current injector due to its highly insulating properties and high Curie temperature (550 K).\cite{glass} In addition, its low magnetic loss properties at microwave frequencies make it ideal for efficient spin injection. The magnetization damping in YIG is two orders of magnitude lower than that in ferromagnetic metals.\cite{sparks}

Among the various fabrication methods of YIG, liquid-phase epitaxy (LPE) is known for its ability to produce high-quality single-crystal films\cite{levinstein, glass} which have been used extensively in spintronics experiments.\cite{kajiwara, uchidanmat2010, kurebayashi} 
However, it is difficult to produce films thinner than a few hundred nanometers by the LPE method.\cite{castel}
Since it has been shown that the interface damping due to SP increases with decreasing thickness of the ferromagnetic film,\cite{tserkovnyak-PRB, tserkovnyak-PRL} synthesis of YIG films with thickness below 100 nm is desirable for the study of interface effects.
Conversely, the increase and saturation of the spin Seebeck signal with increasing YIG thickness has been interpreted as evidence that SSE originates in bulk magnonic spin currents. \cite{kehlberger} Therefore, thin YIG films are also useful for probing the physics of the spin Seebeck effect.

In quest of controlling YIG thickness at nanometer scale, the growth of thin films by pulsed-laser deposition (PLD)\cite{krockenberger, sun, althammer, kelly, sun-wu} and sputtering\cite{okamura, jang, jang2, kang0, kang, mwu, boudiar, yamamoto, wang1, wang2, wang3, wang4, wang5, wang6, marmion} has attracted interest. 
In the sputtering method, the growth of crystals can be realized either by direct epitaxial growth via sputtering at high temperatures \cite{okamura, jang2, wang1, mwu} or by sputtering at room temperature and subsequent post-annealing. \cite{jang, yamamoto, jang2, boudiar, kang0, kang, mwu} Direct epitaxial growth at high temperature can provide crystals of excellent quality.\cite{wang1} However, the sputtering rates are usually very low \cite{jang, kang0} and the sample quality sensitive to the conditions during deposition. In contrast, sputtering at ambient temperature is technologically more accessible as it does not require a high process temperature and enables faster deposition.\cite{jang}

Although there are various industrial advantages to the sputtering method, such as high compatibility with the semiconductors technology, suitability for coating of large areas, and dryness of the preparation process, there are only a limited number of reports on the use of sputtered YIG films in spintronics experiments. \cite{wang1, wang2, wang3, wang4, wang5, wang6, marmion} In this work, we grow thin YIG films by sputtering and subsequent post-annealing and confirm epitaxial growth by transmission electron microscopy (TEM). 
By measuring SP and SSE, we demonstrate that the obtained YIG films are an efficient spin current generator comparable to LPE films.

\section{Methods}

YIG films were deposited by on-axis magnetron rf sputtering on gadolinium gallium garnet (111) (Gd$_3$Ga$_5$O$_{12}$, GGG) substrates with a thickness of 500 $\mu$m. The choice of substrate was due to the close match of the lattice constants and of the thermal expansion coefficients of GGG and YIG. \cite{boudiar} The sputtering target had a nominal composition of Y$_{3}$Fe$_{5}$O$_{12}$. The base pressure was $2.3\times 10^{-5}$ Pa. The substrate remained at ambient temperature during sputtering. The pressure of the pure argon atmosphere was $1.3$ Pa.  The deposition rate was fairly high at $2.7$ nm/min with a sputtering power of 100 W. The as-deposited films were non-magnetic; according to Refs. \citenum{jang, yamamoto, jang2, okamura, boudiar, kang0, kang} such films are amorphous. Crystalization was realized by post-annealing in air at $850$ $^{\circ}$C for 24 hours.  In this study, we focus on films with a thickness of 83 and 96 nanometers. The thickness was determined by X-ray reflection (XRR) and TEM. The structure of the samples was characterized by high-resolution TEM. X-ray photoelectron spectroscopy confirmed a Y:Fe stoichiometry 3:4.4. 

Microwave properties were analyzed using a 9.45-GHz TE$_{011}$ cylindrical microwave cavity and a  coplanar transmission waveguide in the 3-10 GHz range.  The waveguide had a 2-mm-wide signal line and was designed to a 50-$\Omega$ impedance. The width and length of the samples were $w=1$ mm and $l=3$ mm, respectively.

For the spin injection experiments, the annealed YIG samples were coated by a platinum film by rf sputtering. Spin current injected into the platinum layer was detected electrically using the inverse spin Hall effect [ISHE, Fig. \ref{fig1}(a)]. ISHE originates in the spin-orbit interaction which bends the trajectories of electrons with opposite spins and opposite velocities in the same direction and produces an electric field transverse to the direction of  the spin current. \cite{saitoh, costache, azevedo, valenzuela}
Platinum was chosen for its high conversion efficiency from spin current to charge current.\cite{ando}

Spin pumping was performed at room temperature in a cylindrical 9.45-GHz TE$_{011}$ cavity at a microwave power $P_{\text{MW}}=1$ mW (corresponding to a microwave field $\mu_0h_{\text ac}=0.01$ mT) in a setup illustrated in Fig. \ref{fig1}(a). The sample was placed in the centre of the cavity where the electric field component of the microwave is minimized while the magnetic field component is maximized and lies in the plane of the sample surface. A static magnetic field was applied perpendicular to the direction of the microwave field and to the direction in which the voltage was measured.\cite{ando} Measurements were performed on a set of three samples. The thickness of the Pt layer was $d_{N}=14$ nm, the thickness of the YIG layer $d_{F}=96$ nm.

The SSE experiment was performed in a longitudinal setup identical to that of Ref. \citenum{kikkawa} on three YIG(83 nm)$\mid$Pt(10 nm) samples. The length, the width, and the thickness of the samples were $L_V=6$ mm, $w=1$ mm, and $L_T=0.5$ mm, respectively.  
The sample was sandwiched between two insulating AlN plates with high thermal conductivity. 
The upper AlN plate (on top of the Pt layer) was thermally connected to a Cu block held at room temperature. 
The bottom AlN plate (under the GGG substrate) was placed on a Peltier module.
The width of the upper AlN plate (5 mm) was slightly shorter than the sample length (6 mm) in order to take electrical contacts with tungsten needles. The samples were placed in a $10^{-2}$ Pa vacuum in order to prevent heat exchange with the surrounding air. A static magnetic field was applied in the plane of the sample surface perpendicular to the direction in which the voltage was measured.

\section{Results and discussion}

\subsection{Structural and microwave properties}

\begin{figure*}[t]
\floatbox[{\capbeside\thisfloatsetup{capbesideposition={right,center},capbesidewidth=4.5cm}}]{figure}[\FBwidth]{\caption{\label{fig2}
Structural and microwave properties of the YIG films. (a) TEM image of the GGG$\mid$YIG interface. (b) Selected area diffraction at the GGG$\mid$YIG interface with the electron beam along the [0$\overline1$1] axis. (c) The cross section of a GGG$\mid$YIG$\mid$Pt sample. (d) Magnified view of the spherical defects in the YIG structure. (e) Magnified view of the YIG$\mid$Pt interface. (f) In-plane FMR derivative absorption spectrum measured in a microwave cavity. The fit is a derivative of the Lorentzian function. (d) Frequency scan of the FMR peak-to-peak linewidth $W$ measured using a coplanar waveguide (circles) and values obtained in a microwave cavity (squares). The fitting function is given by Eq. \eqref{linewidth}.
}}
{\includegraphics[width=12.1cm, angle=0, bb=0 0 340 290]{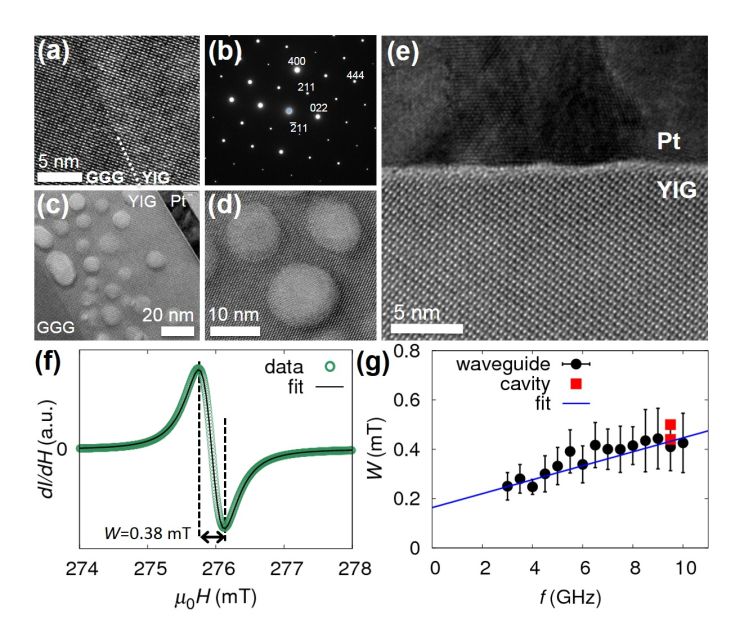}}
\end{figure*}

Figures \ref{fig2}(a)-(e) present the structural properties of the 96-nm-thick films observed by TEM. A magnified view of the GGG$\mid$YIG interface and the diffraction pattern at this interface are shown in Figs. \ref{fig2}(a) and \ref{fig2}(b), respectively. The YIG grows epitaxially on the GGG substrate. Neither defects nor misalignment in the lattice planes were observed in the TEM images [Fig. \ref{fig2}(a)]. As shown in Fig. \ref{fig2}(b),  the diffraction pattern consists of a single reciprocal lattice confirming perfect alignment of the GGG and YIG structures.

An image of the whole cross section of a GGG$\mid$YIG$\mid$Pt sample is shown in Fig. \ref{fig2}(c). The YIG film contains spherical defects with a diameter of roughly 10 nm.  
However, these are suppressed in the vicinity of the YIG$\mid$Pt interface.  
TEM imaging of as-deposited films revealed a uniform amorphous Y-Fe-O layer indicating that the spherical structures emerge during post-annealing.
A magnified view of these objects is given in Fig. \ref{fig2}(d).
They do not possess crystalline structure. This can be also inferred from the fact that only a single reciprocal lattice, corresponding to epitaxial growth, was observed in the diffraction pattern.  
We speculate that these defects are voids which appear due to the volume change in the transition from amorphous to crystalline phase.
There is a possibility that these structures contain residual amorphous material left over in the crystallization.
We expect that these structures, due to their location inside the film, do not affect spin injection efficiency because spin injection originates in the spin transport at the F$\mid$N interface.\cite{tserkovnyak-PRB, tserkovnyak-PRL}

The YIG$\mid$Pt interface is magnified in Fig. \ref{fig2}(e). One can see that the YIG maintains its crystal structure up to the top of the layer. The surface of the YIG film is flat with a roughness less than 1 nm.  This clean interface is of advantage for efficient spin injection. \cite{qiu}

XRR measurement on a bare YIG film yielded a YIG surface roughness of $(0.008\pm 0.002)$ nm. In contrast, the GGG$\mid$YIG interface roughness was $(0.6\pm0.1)$ nm. The fact that the roughness at the GGG$\mid$YIG interface was many times larger than that at the YIG surface can be ascribed to substrate damage caused by on-axis sputtering.

Figure \ref{fig2}(f) shows the ferromagnetic resonance (FMR) derivative absorption spectrum $dI/dH$ of a 96-nm-thick YIG film measured in a microwave cavity at $P_{\text{MW}}=1$ mW. It consists of a single Lorentzian peak derivative with a peak-to-peak linewidth $W=0.38$ mT. This corresponds to a single FMR mode with damping proportional to the linewidth. The linewidth in a broad set of samples varied in the range of $0.4-0.6$ mT. These values are among the lowest reported on sputtered YIG films.\cite{kang, wang1, mwu} The effective saturation magnetization $M_{\text{eff}}$ was determined from the dependence of the FMR field on the direction of the static magnetic field with respect to the sample plane.\cite{ando} The obtained value $M_{\text{eff}}=(103\pm4)$  kA/m is lower than the saturation magnetization value for bulk YIG crystal (140 kA/m).\cite{spin-waves} The decrease in the saturation magnetization might be a result of a deficiency in Fe atoms indicated by the off-stoichiometry.

Figure \ref{fig2}(g) shows the frequency $f$ dependence of the peak-to-peak linewidth $W$ measured on a 83-nm-thick YIG film using a coplanar waveguide. Using a linear fit \cite{spin-waves}
\begin{equation}
W=W_0+\frac{4\pi}{\sqrt{3}}\frac{\alpha}{\gamma}f \label{linewidth}
\end{equation}
with gyromagnetic ratio $\gamma=1.78\times10^{11}$ T$^{-1}$s$^{-1}$ determined from the frequency dependence of the FMR field,\cite{iguchi} we obtain a damping constant  $\alpha=(7.0\pm1.0)\times 10^{-4}$. This value is more than ten times larger than the value for bulk single crystals ($3\times 10^{-5}$),\cite{sparks} but is slightly smaller than other values reported on films prepared by sputtering \cite{wang4, mwu} and only three times higher than values reported on LPE films.\cite{castel, pirro} The increase in the damping constant is probably due to two-magnon scattering on defects in the film.\cite{arias}

\subsection{Spin pumping and spin Seebeck effect}

\begin{figure}[t]
\centerline{\includegraphics[width=8.5cm, angle=0, bb=0 0 240 160]{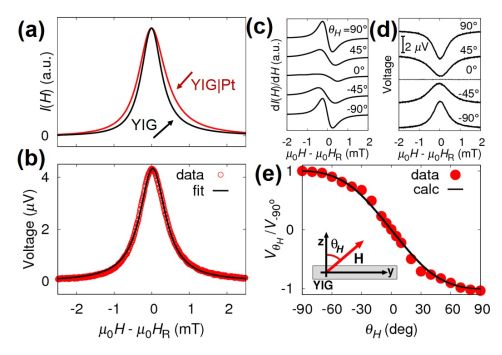}}
\caption{\label{fig3}
Results of the spin pumping measurement on a YIG(96 nm)$\mid$Pt(14 nm) bilayer at $P_{\text{MW}}=1$ mW ($\mu_0 h_{\text{ ac}}=0.01$ mT).  (a) The integrated FMR spectrum of a YIG sample before  and after Pt coating  (YIG and YIG$\mid$Pt, respectively). (b) ISHE voltage signal measured on the Pt layer at $\theta_H=-90^{\circ}$ overlaid with a Lorentzian fit. (c), (d) FMR derivative spectra [(c)] and spectral shapes of the voltage signals [(d)] at selected $\theta_H$ values. (e) Angle dependence of the peak value of the ISHE voltage (``data'') overlaid with the calculated dependence (``calc''). The inset shows the definition of $\theta_H$.
}
\end{figure}

The results of the SP experiment are shown in Fig. \ref{fig3}. Figure \ref{fig3}(a) compares the integrated FMR spectra of the plain YIG film and the YIG$\mid$Pt bilayer measured on one YIG sample prior to and after Pt coating. The linewidth in the YIG$\mid$Pt bilayer increases on average by 30\% as compared to the linewidth in the bare YIG layer. This corresponds to enhanced damping  of the magnetization precession in the YIG$\mid$Pt sample. This enhancement is caused by the transfer of spin angular momentum to conduction electrons in Pt near the YIG$\mid$Pt interface, indicating successful spin injection.

Simultaneously with the FMR peak of the ferromagnet, a voltage signal appears across the Pt layer, as shown in Fig. \ref{fig3}(b). The spectral shape of the voltage signal is a Lorentzian with the same centre and full-width-at-half-maximum as the FMR spectrum of the YIG [Fig. \ref{fig3}(a)]. This is an expected behaviour in ISHE, where the generated voltage at field $H$ is proportional to the microwave absorption intensity $I(H)$.\cite{ando}

The fact that the detected voltage signal is due to ISHE is confirmed by the $\theta_H$ dependence, where  $\theta_H$ is the angle between the surface normal and the magnetic field [see inset of Fig. \ref{fig3}(e)].
The FMR derivative spectra and the spectral shapes of the voltage signals at selected values of $\theta_H$ are shown in Figs. \ref{fig3}(c) and (d), respectively.  The spectral shape of the voltage copies the shape of $I(H)$ even when the magnetic field is tilted out of the sample plane ($\theta_H=\pm 45^\circ$).
The sign of the voltage reverses by reversing the direction of the magnetic field and the signal vanishes when the magnetic field is perpendicular to the sample surface ($\theta_H=0^{\circ}$). This is a signature of ISHE, where the electromotive force is generated along the vector product of the spin polarization and the spin current, $\mathbf{E}_{\text{ISHE}}\propto \mathbf{j}_{s} \times \boldsymbol{\sigma}$. \cite{saitoh, ando}

Figure \ref{fig3}(e) shows the $\theta_H$ dependence of the peak value of the voltage signal. The black curve is the expected $\theta_H$ dependence of the ISHE voltage calculated following the procedure in Ref. \citenum{ando}. The magnitude of the ISHE voltage is proportional to the injected spin current, $V_{\text{ISHE}} \propto j_s\sin \theta_M$, where the spin current magnitude $j_s$ is given by Eq. (12) in Ref. \citenum{ando}

\begin{widetext}
\begin{equation}
j_s=\frac{g_r^{\uparrow \downarrow} \gamma^2 (\mu_0 h_{\text{ac}})^2 \hbar
\left[
{\mu_0 M_{\text{eff}}} \gamma \sin^2 \theta_M + \sqrt{({\mu_0 M_{\text{eff}}})^2 \gamma^2 \sin^4 \theta_M + 4 \omega^2}
\right]
}{8\pi\alpha^2
\left[
({\mu_0 M_{\text{eff}}})^2 \gamma^2 \sin^4 \theta_M + 4\omega^2
\right]
}. \label{sc}
\end{equation}
\end{widetext}
Here $\theta_M$ is the angle between the magnetization vector and the surface normal, and $g_r^{\uparrow \downarrow}$ the real part of the spin mixing conductance.  The relation between $\theta_H$ and $\theta_M$ is determined by the resonance condition 
$
\left( \omega/\gamma \right)^2=
\left[ \mu_0 H_{\text{R}} \cos (\theta_H-\theta_M) -{\mu_0 M_{\text{eff}}} \cos 2\theta_M
\right]
\times
\left[
\mu_0 H_{\text{R}} \cos (\theta_H-\theta_M) -{\mu_0 M_{\text{eff}}} \cos^2 \theta_M
\right]
$
 [Eq. (9) in Ref. \citenum{ando}] and the static equilibrium condition 
$
2\mu_0 H\sin(\theta_H-\theta_M)+{\mu_0 M_{\text{eff}}} \sin 2 \theta_M=0
$
[Eq. (6) in Ref. \citenum{ando}].  To numerically calculate Eq. \eqref{sc}, we used $\omega=5.94\times10^{10}$ s$^{-1}$, $\gamma=1.78\times10^{11}$ s$^{-1}$T$^{-1}$ and {$M_{\text{eff}}=103$ kA/m}. The result of the calculation is in very good agreement with the data, providing another piece of evidence that the observed voltage is due to ISHE.

\begin{figure}[b]
\centerline{\includegraphics[width=8.5cm, angle=0, bb=0 0 240 115]{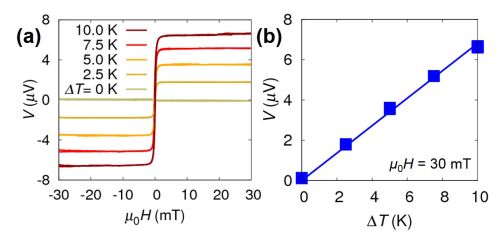}}
\caption{\label{fig4}
Results of the SSE measurement on a YIG(83 nm)$\mid$Pt(10 nm) bilayer. (a) Magnetic field $\mu_0 H$  dependence of the voltage signal $V$ on the Pt layer measured for different values of temperature difference $\Delta T$ across the GGG$\mid$YIG$\mid$Pt sample in the longitudinal SSE configuration. (b) $\Delta T$ dependence of the voltage magnitude at $\mu_0H=30$ mT.
}
\end{figure}

Figure \ref{fig4} shows the results of the longitudinal SSE measurement. Figure \ref{fig4}(a) gives the magnetic field $\mu_0H$ dependence of the voltage signal $V$ measured on the Pt layer for selected values of temperature difference $\Delta T$ between the bottom and the top of the sample. We observed a voltage signal whose sign is reversed by reversing the direction of the magnetic field. Upon increasing the magnetic field, the magnitude of the signal increases monotonically until reaching a saturation value. This $\mu_0H$ dependence of $V$ reflects the magnetization curve of YIG.\cite{kikkawa} The saturation value of the voltage increases with increasing temperature gradient. No signal was observed for $\Delta T=0$ K. As shown in Fig. \ref{fig4}(b), the magnitude of the voltage at $\mu_0 H=30$ mT is linear in $\Delta T$. This behaviour is consistent with ISHE induced by SSE, where the spin current generated across the YIG$\mid$Pt interface is proportional to the temperature gradient $\boldsymbol{\nabla} T$.\cite{uchidaAPL}

\subsection{Spin injection efficiency}

The normalized value of the ISHE electromotive force observed in the SP experiment is $E_{\text {ISHE}}/(\mu_0 h_{\text {ac}})=(150\pm30)$ $\mu$V/(mm$\cdot$mT). This is a few times higher than the value reported on a 4.5 $\mu$m-thick LPE film, $E_{\text {ISHE}}/(\mu_0 h_{\text {ac}})=39$ $\mu$V/(mm$\cdot$mT), measured at the same equipment, \cite{qiu} and comparable with values reported for LPE films of 1.2-$\mu$m thickness [160 $\mu$V/(mm$\cdot$mT) in Ref. \citenum{kajiwara}].

As for the ISHE voltage in the SSE measurement, using the experimental value $V=(5.6 \pm 1.2)$ $\mu$V at $\Delta T=10$ K, we obtain a normalized voltage $V\times L_T/L_V =(0.47\pm0.10)$ $\mu$V. This is also of the same order as the value for YIG prepared by LPE (1 $\mu$V in Ref. \citenum{kikkawa} for a 4.5-$\mu$m-thick film).

Finally, we estimate the spin mixing conductance at the YIG$\mid$Pt interface. The efficiency of the transfer of spin angular momentum at the F$\mid$N interface is described by the real part of the spin-mixing conductance $g_r^{\uparrow \downarrow}$,\cite{tserkovnyak, kajiwara, xia, jia} which is given by \cite{tserkovnyak, mosendz, burrowes}

\begin{equation}
g_r^{\uparrow \downarrow}=
\frac{4\pi {M_{\text{s}}} d_F}{g \mu_{\rm B}} \frac{\sqrt{3}\gamma}{2\omega} \left( W_{F/N}-W_{F}
\right). \label{gr}
\end{equation}
Here, $g$ is the $g$-factor, $\mu_{\rm B}= e \hbar/(2m_e)=9.27\times10^{-24}$ J$\cdot$T$^{-1}$ the Bohr magneton, $M_\text{s}$ the saturation magnetization; and $W_F$ and $W_{F/N}$ are the peak-to-peak linewidth of the FMR spectrum in the bare ferromagnetic film and in the F$\mid$N bilayer, respectively. Using $g=2.12$, $M_\text{s}\approx M_{\text{eff}}=103$ kA/m, $\omega/\gamma=0.334$ T, $d_F=96$ nm, $W_F=(0.40\pm0.03)$ mT and $W_{F/N}=(0.52\pm0.02)$ mT, we obtain a spin-mixing conductance $g_r^{\uparrow \downarrow}=(2.0\pm 0.2)\times10^{18}$ m$^{-2}$. 
This value is of the same order as those in other reports on the YIG$\mid$Pt interface, e.g. $g_r^{\uparrow \downarrow}=1.3\times10^{18}$ m$^{-2}$ in Ref. \citenum{qiu}.  
Thus, both in SP and in SSE, we have obtained spin injection efficiencies comparable to those reported at LPE-made-YIG$\mid$Pt bilayer samples. 
This result suggests that defects inside the YIG film do not significantly affect the transfer of spin angular momentum at the interface with Pt. The high spin injection efficiency is promoted by the presence of a regular garnet structure in the vicinity of the interface with Pt as well as the clean interface, as observed by TEM imaging.

It is worth noting that
based on Eqs. \eqref{sc} and \eqref{gr}, a decrease in magnetization from 140 kA/m to 103 kA/m should lead to a 28 \% decrease in the injected spin current. A corresponding suppression of the inverse spin Hall voltage should be observed. However, the errors in the spin pumping and the SSE voltage measurements were 20 \% and 21 \%, respectively. This degree of error does not allow to discuss the effect of the decreased magnetization on spin pumping.

To conclude, we estimate the spin Hall angle of Pt from the obtained data. The injected spin current determined from Eq. \eqref{sc} is $j_s=5.3\times10^{-10}$ J/m$^{2}$, where we have used 
$g_{r}^{\uparrow \downarrow}=2.0\times 10^{18}$ m$^{-2}$,
$\mu_0 h_{\text{ac}}=0.01$ mT, $\alpha=7.0\times 10^{-4}$, $\gamma=1.78\times 10^{11}$ T$^{-1}$s$^{-1}$, $\omega=5.94\times 10^{10}$ s$^{-1}$, $M_{\text{eff}}=103$ kA/m, and $\theta_H=\theta_M=-90$ $^\circ$.
The peak value of the inverse spin Hall voltage is given by\cite{ando}

\begin{equation}
V_{\text{ISHE}}=\frac{d_E\theta_{\text{SHE}}\lambda_N \tanh(d_N/2\lambda_N)}{d_N\sigma_N}
\left( \frac{2e}{\hbar} \right)j_s,
\end{equation}
where $d_E$ is the distance of the electrodes, $\theta_{\text{SHE}}$ the spin Hall angle, $\lambda_N$ the spin diffusion length in Pt, $d_N$ the thickness of the Pt layer, and $\sigma_N$ the conductivity of Pt.   Using $V_{\text{ISHE}}=3.9$ $\mu$V, $d_E=2.6$ mm, $d_N=14$ nm, $\sigma_N=3.1\times10^6$ $\Omega^{-1}$m$^{-1}$, and $j_s=5.3\times10^{-10}$ J/m$^{2}$, we obtain $\theta_{\text{SHE}}=0.029$ or $0.007$ for Pt spin diffusion length $\lambda_N=1.4$ nm,\cite{lliu} or 10 nm,\cite{vila} respectively. Both values are within the range of spin Hall angles reported for Pt.

\section{Conclusions}

In summary, we have prepared YIG films by the sputtering method and investigated their structural and microwave properties as well as spin current generation from these films.
The results show that the presented fabrication method, consisting of sputtering at room temperature and post-annealing in air, provides epitaxial YIG films with thickness below 100 nm which have excellent microwave properties in spite of defects in the structure. The spin injection efficiency observed in spin pumping and in the spin Seebeck effect is comparable with that for high-quality films prepared by liquid phase epitaxy. The above preparation of garnet films is relatively straightforward and the technological requirements modest. Moreover, this method offers a possibility to control the YIG thickness at the nanometer scale. These results are of potential use in spintronics research.

\begin{acknowledgments}
We thank S. Ito from Analytical Research Core for Advanced Materials, Institute for Materials Research, Tohoku University, for performing transmission electron microscopy on our samples.
This work was supported by 
CREST ``Creation of Nanosystems with Novel Functions through Process Integration'',
 PRESTO ``Phase Interfaces for Highly Efficient Energy Utilization'', 
Strategic International Cooperative Program ASPIMATT from JST, Japan, 
Grant-in-Aid for 
Challenging Exploratory Research (Nos. 26610091 and 26600067), 
Research Activity Start-up (No. 25889003), 
Young Scientists (B) (No. 26790038), 
Young Scientists (A) (No. 25707029), 
and Scientific Research (A) (No. 24244051) from MEXT, Japan,
LC-IMR of Tohoku University, the Tanikawa Fund Promotion of Thermal
Technology, the Casio Science Promotion Foundation, and the Iwatani Naoji
Foundation.
\end{acknowledgments}

\end{document}